\documentclass{article}

\usepackage{spconf}
\usepackage{amsmath}
\usepackage{graphicx}
\usepackage{multicol}
\usepackage{url}
\usepackage{color,booktabs} 
\usepackage{amsfonts} 
\usepackage{xcolor}

\DeclareMathOperator{\elu}{ELU}
\DeclareMathOperator{\sigmoid}{sigmoid}

\title{Timbre-Trap: A Low-Resource Framework for Instrument-Agnostic Music Transcription}

\name{
\begin{tabular}{@{}c@{}}
Frank Cwitkowitz$^{1, \dagger}$\thanks{$\dagger$ Work completed as a research intern at Sony.}
\qquad Kin Wai Cheuk$^{2}$
\qquad Woosung Choi$^{2}$\\
\qquad Marco A. Martínez-Ramírez$^{2}$
\quad Keisuke Toyama$^{3}$
\quad Wei-Hsiang Liao$^{2}$
\quad Yuki Mitsufuji$^{2, 3}$
\end{tabular}}
\address{$^{1}$University of Rochester
         \quad $^{2}$Sony AI
         \quad $^{3}$Sony Group Corporation}

\usepackage{tikz}
\newcommand\copyrightnote{
\begin{tikzpicture}[remember picture, overlay]
\node[anchor=south, yshift=15pt] at (current page.south) {\fbox{\parbox{\dimexpr\textwidth - \fboxsep - \fboxrule\relax}
{\scriptsize Copyright 2024 IEEE. Published in ICASSP 2024 - 2024 IEEE International Conference on Acoustics, Speech and Signal Processing (ICASSP), scheduled for 14-19 April 2024 in Seoul, Korea. Personal use of this material is permitted. However, permission to reprint/republish this material for advertising or promotional purposes or for creating new collective works for resale or redistribution to servers or lists, or to reuse any copyrighted component of this work in other works, must be obtained from the IEEE. Contact: Manager, Copyrights and Permissions / IEEE Service Center / 445 Hoes Lane / P.O. Box 1331 / Piscataway, NJ 08855-1331, USA. Telephone: + Intl. 908-562-3966.}}};
\end{tikzpicture}}

\begin{document}
\ninept
\maketitle
\begin{abstract} 
In recent years, research on music transcription has focused mainly on architecture design and instrument-specific data acquisition. With the lack of availability of diverse datasets, progress is often limited to solo-instrument tasks such as piano transcription. Several works have explored multi-instrument transcription as a means to bolster the performance of models on low-resource tasks, but these methods face the same data availability issues. We propose Timbre-Trap, a novel framework which unifies music transcription and audio reconstruction by exploiting the strong separability between pitch and timbre. We train a single autoencoder to simultaneously estimate pitch salience and reconstruct complex spectral coefficients, selecting between either output during the decoding stage via a simple switch mechanism. In this way, the model learns to produce coefficients corresponding to timbre-less audio, which can be interpreted as pitch salience. We demonstrate that the framework leads to performance comparable to state-of-the-art instrument-agnostic transcription methods, while only requiring a small amount of annotated data.
\end{abstract}

\section{Introduction}
\label{sec:introduction}
\copyrightnote
Automatic music transcription (AMT) is a heavily researched and important topic within the Music Information Retrieval (MIR) field \cite{benetos2018automatic}. Generally, the task involves estimating the fundamental frequencies (F0s) of overlapping pitched events within audio signals containing music. Pitch characterizes notes, the building blocks of music, as well as their articulation, and is extremely relevant for many types of musical analysis. AMT has the potential to enable the next generation of interactive music systems, as well as the potential to produce meaningful low-level features for other MIR tasks.

There are several distinctions one should make when characterizing AMT systems, including frame- vs. note-level, instrument-informed vs. instrument-agnostic, and single- vs. multi-instrument \cite{wu2020multi}. While monophonic AMT approaches are very mature \cite{morrison2023cross}, there are still serious challenges to overcome when analyzing polyphonic music. Polyphony can arise from either polyphonic instruments or multi-instrument mixtures. In order to simplify the problem, it is very common to design AMT systems for solo-instrument tasks. This assumption makes it easier to learn and exploit domain-specific characteristics, but such models have limited applications, require large amounts of instrument-specific training data, and neglect the broader similarities and relationships among various AMT tasks.

In this work, we address the problem of frame-level, instrument-agnostic, multi-instrument AMT, which is also known as Multi-Pitch Estimation (MPE). Here, the idea is to estimate the high-resolution pitch activity of concurrently-played musical instruments within each frame of an audio signal, without making an effort to classify or assign pitch activity to any particular instrument class or instance. We argue that musical audio can be loosely decomposed into pitch activity and timbre, and as such formulate MPE as a guided reconstruction problem. Specifically, we train an autoencoder to estimate pitch activity or salience $Y \in [0, 1]^{K\times T}$ while simultaneously reconstructing complex Constant-Q Transform (CQT) \cite{velasco2011constructing} spectral coefficients $X \in \mathbb{C}^{K\times T}$. Since the vertical ($K$) and horizontal ($T$) axis of $Y$ and $X$ have exactly the same meaning, it is quite natural to switch between them. As such, we incorporate a simple switch mechanism into the latent space of our model to select between reconstruction and transcription. In this way, transcription can be thought of as filtering out timbre information from the audio.

The process of acquiring pitch or note annotations for musical audio is challenging, expensive, time-consuming, and error-prone. This has resulted in very limited training data for many AMT tasks. With the proposed AMT formulation, the reconstruction objective can guide a model to extract more out of available training data. This makes our approach an attractive means to address low-resource AMT tasks, such as those involving less common instruments or instruments for which it is particularly challenging to produce annotations. We conduct experiments to demonstrate that our model can perform MPE effectively with limited training data, and that our approach is suitable for low-resource transcription tasks. We also conduct an ablation study to analyze the influence of the proposed training objectives and to verify several architectural design choices.

\section{Background \& Related Work}
\label{sec:background_related_work}
Traditionally, AMT systems have employed signal processing \cite{klapuri2007signal} or spectral decomposition \cite{benetos2015efficient} techniques to analyze audio in the frequency domain. While these methods can perform well, they do not tend not to generalize to out-of-domain audio and typically involve algorithmic inputs that require careful tuning. At present, AMT is largely dominated by deep neural networks (DNNs) which map time-frequency features to a pianoroll-like or pitch salience representation. These output representations characterize activity for a discrete set of non-mutually exclusive pitch classes over time. Common choices of time-frequency features include the Short-Time Fourier Transform (STFT), which is lossless and has constant frequency resolution, and the Mel Spectrogram or CQT, which more closely resemble the logarithmic frequency perception of human hearing. Early DNNs have comprised mainly convolutional and recurrent layers \cite{sigtia2016end, bittner2017deep, hawthorne2018enabling}, whereas some more recent models leverage self-attention or adopt transformer-style architectures \cite{wu2020multi, gardner2021mt3, lu2023multitrack}.

Another class of AMT models employ autoencoder architectures to better exploit the strong relationship between input time-frequency features and the desired output pitch salience. Autoencoders have been applied to learn more powerful features \cite{pedersoli2020improving, cheuk2021reconvat, weiss2022comparing, wu2023mfae} for AMT models, as well as to perform pitch salience prediction directly \cite{wu2020multi}. Although there are existing methods which leverage spectrogram reconstruction to improve transcription accuracy \cite{cheuk2021effect, cheuk2021reconvat}, these require a separate model to perform reconstruction given pitch information. In this work, we attempt to unify both transcription and reconstruction using a single model. Music audio can be loosely decomposed into pitch activity and timbre. As timbre is filtered out of an audio signal, the remainder of the signal approaches a superposition of pure sinusoidal tones. As such, we argue that transcription and reconstruction are intrinsically related objectives which can benefit from each other. A similar assumption is made for Differentiable Digital Signal Processing (DDSP) models \cite{engel2020ddsp, engel2020self}, however these models are typically monophonic and have no mechanism for supervising transcription. In contrast, Timbre-Trap applies a supervised transcription objective similar to existing AMT methods, while simultaneously applying a self-supervised reconstruction objective.

\section{Proposed Method}
\label{sec:proposed_method}

\subsection{Complex CQT Features}
\label{sec:complex_CQT_features}
A common practice for AMT is to use the CQT to compute audio features, due to its ability to represent spectral energy using a set of bases with geometrically spaced center frequencies and a constant frequency-to-bandwidth ratio or Q-factor \cite{brown1991calculation}. These properties are very useful for musical analysis, since human perception of pitch is geometric and frequency bins can be mapped directly to notes within the Western music scale. However, the CQT by default is not invertible. While typically irrelevant for AMT, invertible CQT formulations have been applied to audio inverse problems to enable high-quality musical audio synthesis \cite{moliner2023solving}. Since our framework leverages audio reconstruction, we employ a slice-based implementation\footnote{Found at \url{https://github.com/archinetai/cqt-pytorch}.} of the invertible non-Stationary Gabor Transform (NSGT) CQT formulation proposed in \cite{velasco2011constructing, holighaus2012framework}. We compute complex CQT spectral coefficients $X \in \mathbb{C}^{K\times T}$ spanning $9$ octaves below the Nyquist frequency with $60$ bins per octave at an effective hop size of $2.93$ ms, processing $3$ second slices of audio resampled to $22050$ Hz at a time. As in \cite{moliner2023solving}, complex coefficients are converted to dual-channel feature maps representing the corresponding real and imaginary parts. Besides encouraging better reconstruction, this formulation helps establish a clear connection between reconstruction and transcription, described in Sec. \ref{sec:model_architecture}, and allows the model to account for phase information when performing transcription. One byproduct of this formulation is the ability to synthesize reconstructed or transcribed coefficients, allowing one to subjectively analyze the quality of either through listening or to remove timbre from an audio signal.

\subsection{Model Architecture}
\label{sec:model_architecture}
In this work we employ a fully convolutional autoencoder, taking inspiration from the architecture of the SoundStream audio compression model \cite{zeghidour2021soundstream}. Our model receives and produces dual-channel time-frequency features representing the real and imaginary parts to complex CQT coefficients. The encoder consists of an initial 2D convolutional layer which doubles the number of channels, four encoder blocks, each of which double the number of channels and halve features along the vertical axis, and a final 1D convolution along frequency to reduce features to latent vectors of size $128$. Each encoder block comprises $3$ residual convolutional blocks with dilation 1, 2, and 3, respectively, followed by a strided 1D convolution which doubles the number of channels and has kernel size $4$ and stride $2$ to reduce dimensionality along the vertical axis. The residual blocks each consist of a dilated 2D convolutional layer with kernel size 3 and a $1\times1$ convolutional layer. The decoder follows exactly the reverse structure of the encoder, but substitutes the latent space convolution and strided convolutions with transposed convolutions. All convolutional layers besides the final layer of the encoder and decoder are followed by $\elu$ activation.

An additional binary channel indicating transcription vs. reconstruction is inserted at the latent space in order to condition the decoder to produce each respective output. When reconstruction mode is selected, the dual-channel output $X'$ is interpreted directly as the real and imaginary parts to complex CQT coefficients. However, when transcription mode is selected, the magnitude of the output coefficients is taken and compressed $Y' = tanh(|X'|)$ using $\tanh$ activation. The subsequent values are then interpreted as pitch salience predictions, which follow a probability-like distribution. Without the complex CQT module, such a connection between magnitude spectral coefficients and pitch salience may be weaker.

\begin{figure}[t]
\centering
\includegraphics[width=0.95\columnwidth]{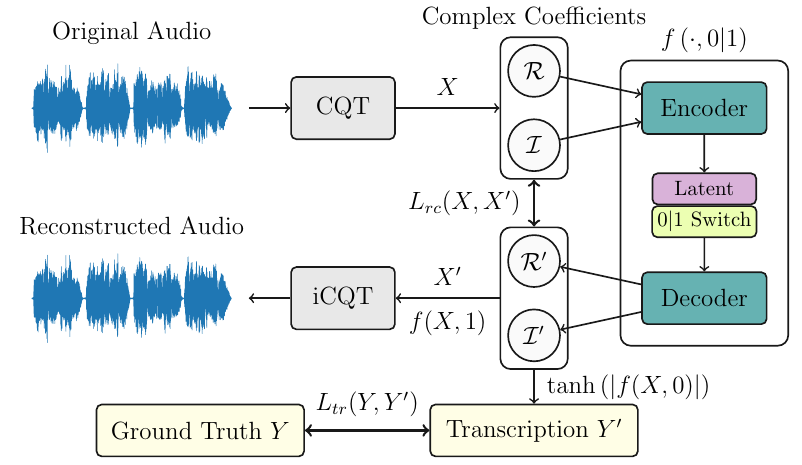}
\caption{Proposed reconstruction-guided transcription framework. Audio is transformed into complex CQT coefficients, which are fed into an autoencoder to produce either reconstructed CQT coefficients or an estimated pitch salience, based on a binary switch.}
\label{fig:framework}
\end{figure}

\subsection{Training Objectives}
\label{sec:training_objectives}
Our model is trained with two main objectives, transcription and reconstruction, and an auxiliary consistency objective, all of which are formulated using mean-squared error (MSE) loss. Given model $f(\cdot, 0|1)$, complex CQT coefficients $X$, and binary ground-truth $Y$, we can estimate the pitch salience $Y' = tanh(|f(X, 0)|)$ and spectral coefficients $X' = f(X, 1)$. Transcription loss is defined as 
\begin{equation}
\label{eq:transcription_loss}
L_{tr}(Y, Y') = \frac{1}{T} \sum_{T} \sum_{K} \gamma \times \left( Y - Y' \right)^2,
\end{equation}
where $\gamma = Y \times \frac{\sum \neg Y}{\sum Y} + (1 - Y)$ is a scaling tensor to balance the loss for positive and negative salience targets. Minimizing this objective is analogous to minimizing binary cross-entropy loss under standard AMT frameworks. Similarly, reconstruction loss is defined as
\begin{equation}
\label{eq:reconstruction_loss}
L_{rc}(X, X') = \frac{1}{T} \sum_{T} \sum_{K} \mathcal{R} \left( X - X' \right)^2 + \mathcal{I} \left( X - X' \right)^2.
\end{equation}
Under the Timbre-Trap framework, $L_{rc}$ behaves as a regularizer for MPE and encourages our model to discover relationships between pitch and timbre that are useful for transcription. Finally, we introduce an auxiliary consistency loss, defined as
\begin{equation}
\label{eq:consistency_loss}
L_{cn} = L_{rc}(Z, f(Z, 0)) + L_{rc}(Z, f(Z, 1)),
\end{equation}
where $Z = f(X, 0)$ are complex coefficients produced when transcription mode is selected. Assuming transcription coefficients represent timbre-less audio, the model should produce identical coefficients for both transcription and reconstruction when transcription coefficients are fed to the model as input. This loss encourages the model to produce transcription coefficients interpretable as audio, while discouraging the model from filtering out any information unrelated to timbre. In other words, when transcription coefficients, or timbre-less audio, is given as input, the original pitch salience should be preserved for transcription and the coefficients should be reconstructed as if they were normal audio for reconstruction. This behavior is important to ensure the two objectives do not conflict with one another. With all three terms, we can compute total loss as
\begin{equation}
\label{eq:total_loss}
L_{tot} = L_{tr} + L_{rc} + L_{cn}.
\end{equation}

\section{Experiments}
\label{sec:experiments}
We train the model described in Sec. \ref{sec:model_architecture} under several different configurations using our proposed framework. In each experiment, the URMP \cite{li2018creating} dataset is used for training and validation following the splits proposed in \cite{gardner2021mt3}. The Bach10 \cite{duan2010multiple} and Su \cite{su2016escaping} datasets are used for evaluation, along with the player $05$ split of GuitarSet \cite{quingyang2019guitarset} to test out-of-domain generalization. All training and evaluation datasets used are polyphonic. Results for Deep Salience \cite{bittner2017deep} and Basic Pitch \cite{bittner2022lightweight}, two baseline MPE models, are included for reference\footnote{We cannot make direct comparisons to these baselines. There is only partial domain overlap between the data used to train them and URMP. Note that a majority of GuitarSet was included in the training set for Basic Pitch.}.
Deep Salience was trained on roughly 10 hours of multitrack data from MedleyDB \cite{bittner2014medleydb}. Basic Pitch was trained on several datasets, including a large collection of synthetic multitracks, a large polyphonic piano dataset, stems from MedleyDB, GuitarSet, and vocal data. Salience thresholds 0.3 and 0.27 were used to acquire pitch estimates for Deep Salience and Basic Pitch, respectively.

\subsection{Evaluation Metrics \& Training Details}
\label{sec:evaluation_metrics_training_details}
In order to assess model performance, we compute frame-level precision ($P$), recall ($R$), and $F_1$-score ($F_1$) for multi-pitch estimates using the conventional \texttt{mir\_eval} package \cite{raffel2014mir_eval}. During inference, we apply local peak-picking and threshold at $0.5$ to generate estimates from pitch salience. Since our model estimates complex CQT coefficients, Signal-to-Distortion Ratio (SDR) is computed as an objective measurement of reconstruction quality. Final scores are obtained by averaging across all tracks within an evaluation set.

Each model is trained on batches of $9$ second excerpts using AdamW optimizer with learning rate $1$E$-3$ and a batch size of $8$. Learning rate warmup is applied over the first $50$ epochs of training, and the learning rate is halved whenever $F_1$-score has not improved for $500$ epochs, with a cooldown time of $100$ epochs. Gradient clipping with an $L_2$-norm of $10$ is applied to improve training stability. The final model for each experiment is chosen as the checkpoint with the maximum $F_1$-score on the validation set across $5000$ epochs. Note that during an epoch only one excerpt per track is sampled.

\subsection{Ablations \& Augmentations}
\label{sec:ablations_augmentations}
In order to investigate the influence of various framework and architectural design choices, experiments for several ablations and augmentations are conducted in addition to the base model experiment. The first set of ablations involves removing different terms from (\ref{eq:total_loss}) to verify whether each term assists or hinders the overall transcription objective. As such, experiments are carried out where $L_{tot}' = L_{tr} + L_{rc}$, \textit{i.e.} no consistency objective, and $L_{tot}'' = L_{tr}$, \textit{i.e.} only the transcription objective, are used to train the model.

The next set of ablations replaces the dual-channel complex CQT coefficients described in Sec. \ref{sec:complex_CQT_features} with their equivalent single-channel magnitude representation in normalized decibels. Since the output of the model then becomes single-channel logits, $\sigmoid$ activation is used to convert to pitch salience and normalized decibels for transcription and reconstruction, respectively. Under this configuration, Timbre-Trap is nearly identical to the standard AMT framework, however it is no longer intuitive to apply the consistency objective, since the transcription coefficients may no longer resemble audio. As such, experiments are carried out with the same two re-formulations of $L_{tot}$ as in the previous set of ablations, \textit{i.e.} no consistency objective and only the transcription objective.

Another set of experiments seeks to investigate whether transcription can improve as a result of simple architectural modifications. When designing a U-Net model, it is common to include skip-connections between corresponding encoder and decoder blocks. Skip-connections can improve gradient flow and prevent the loss of information when downsampling embeddings. We conduct one experiment using weighted skip-connections with learnable weights. Since the decoding stage of Timbre-Trap is conditioned, we also experiment with inserting a Feature-wise Linear Modulation (FiLM) conditioning layer \cite{perez2018film} before the decoder, rather than dedicating an additional latent space channel to indicate the function to perform.

Finally, since $L_{rc}$ is self-supervised, we conduct several experiments to study the effect of training on unlabeled audio in parallel. Only $L_{rc}$ is computed for unlabeled audio, and the batch size is increased to $16$ for these experiments in order to ensure that the effective batch size for labeled data is held constant with respect to all other experiments. In separate experiments, training is carried out with audio-only data from splits $00$--$04$ of GuitarSet, MedleyDB \cite{bittner2014medleydb, bittner2016medleydb}, and the FMA \cite{defferrard2016fma} small subset.

\begin{table*}[t]
\footnotesize
\begin{center}
  \begin{tabular}{|c||c|c|c|c||c|c|c|c||c|c|c|c|}
    \hline
    & \multicolumn{4}{c||}{\textbf{Bach10}} & \multicolumn{4}{c||}{\textbf{Su}} & \multicolumn{4}{c|}{\textbf{GuitarSet}} \\
    \hline
    \multicolumn{1}{|c||}{\textbf{Experiment}} & $\mathit{P}$ & $\mathit{R}$ & $\mathit{F_1}$ & SDR & $\mathit{P}$ & $\mathit{R}$ & $\mathit{F_1}$ & SDR & $\mathit{P}$ & $\mathit{R}$ & $\mathit{F_1}$ & SDR \\
    \hline
    \hline
    Deep Salience \cite{bittner2017deep} & $86.0$ & $61.0$ & $71.3$ & - & $\mathbf{74.1}$ & $47.8$ & $\mathbf{57.1}$ & - & $\mathbf{74.8}$ & $74.0$ & $\mathbf{72.7}$ & - \\
    \hline
    Basic Pitch \cite{bittner2022lightweight} & $\mathbf{86.6}$ & $79.6$ & $\mathbf{82.9}$ & - & $51.5$ & $48.5$ & $48.9$ & - & \textcolor{gray}{$72.8$} & \textcolor{gray}{$81.1$} & \textcolor{gray}{$76.2$} & - \\
    \hline
    Timbre-Trap (ours) & $81.2$ & $84.2$ & $82.6$ & $7.4$ & $52.0$ & $53.0$ & $51.4$ & $5.0$ & $50.1$ & $77.7$ & $60.2$ & $4.0$ \\
    \hline
    \hline
    No Consistency & $78.6$ & $85.0$ & $81.6$ & $8.5$ & $49.9$ & $54.2$ & $50.8$ & $5.8$ & $48.3$ & $77.3$ & $58.4$ & $4.4$ \\
    \hline
    Transcription Only & $78.9$ & $80.8$ & $79.8$ & - 
    & $50.8$ & $47.8$ & $47.9$ & - 
    & $45.5$ & $74.2$ & $55.3$ & - 
    \\
    \hline
    \hline
    Magnitude No Cn. & $79.3$ & $84.0$ & $81.5$ & - & $50.7$ & $54.3$ & $50.6$ & - & $47.6$ & $78.2$ & $58.1$ & - \\
    \hline
    Magnitude Tr. Only & $79.6$ & $80.9$ & $80.2$ & - & $53.3$ & $49.3$ & $49.6$ & - & $53.2$ & $76.2$ & $61.8$ & - \\
    \hline
    \hline
    w/ Skip-Connections & $78.4$ & $81.2$ & $79.7$ & $\mathbf{24.8}$ & $51.4$ & $49.8$ & $49.4$ & $\mathbf{22.0}$ & $47.2$ & $73.7$ & $56.4$ & $\mathbf{18.7}$ \\
    \hline
    w/ FiLM Layers & $80.2$ & $76.5$ & $78.2$ & $7.1$ & $52.2$ & $50.0$ & $49.8$ & $4.1$ & $48.1$ & $73.2$ & $57.0$ & $2.6$ \\
    \hline
    \hline
    + GuitarSet Audio & $80.0$ & $84.7$ & $82.3$ & $6.2$ & $51.6$ & $53.2$ & $50.9$ & $4.6$ & $45.7$ & $79.6$ & $57.4$ & $4.7$ \\
    \hline
    + MedleyDB Audio & $77.2$ & $\mathbf{86.8}$ & $81.7$ & $7.2$ & $49.4$ & $\mathbf{56.0}$ & $51.2$ & $5.5$ & $47.7$ & $80.7$ & $59.2$ & $4.5$ \\
    \hline
    + FMA Audio & $78.3$ & $84.0$ & $81.0$ & $8.4$ & $47.7$ & $54.3$ & $49.7$ & $6.7$ & $44.5$ & $\mathbf{80.7}$ & $56.7$ & $5.7$ \\
    \hline
  \end{tabular}
\end{center}
\caption{Comparison of results for various configurations of the Timbre-Trap framework along with results for reference model reproductions.}
\label{tab:results}
\end{table*}

\subsection{Results}
\label{sec:results}
Results for all baseline methods and experiments are given in Table \ref{tab:results}. We observe that despite training only on $35$ pieces from URMP, which contains $44$ pieces with a total length of 1.3 hours, Timbre-Trap achieves results comparable, and in some cases superior to the reference models. However, it is unable to generalize to GuitarSet as effectively as the reference models. In addition to transcription, Timbre-Trap has the capability of reconstructing audio at a modest SDR or synthesizing the complex coefficients corresponding to pitch salience\footnote{Demos found at \url{https://sony.github.io/timbre-trap}.}. The results indicate that it is possible to model both transcription and reconstruction using a single output representation.

In terms of ablations, we find that the full Timbre-Trap framework with all objectives yields the best overall performance. The consistency and reconstruction objectives both contribute marginally, however we note that experiments modeling complex CQT coefficients without the reconstruction objective exhibited decreased training stability. Although the full framework including the complex CQT formulation allows for the synthesis of audio and pitch salience, the magnitude CQT configuration still leads to comparable performance. Somewhat surprisingly, in some cases the magnitude CQT configuration appears to benefit from the reconstruction objective. This suggests that although magnitude CQT coefficients follow a slightly different distribution than pitch salience, transcription models can potentially benefit from learning to reconstruct them.

The addition of skip-connections leads to a significant increase in reconstruction quality, but hinders overall transcription performance. One potential explanation is that the skip-connections allow timbre information to bypass the bottleneck too easily, undermining the decoder in its effort to filter it out. Another possibility is that skip-connections make reconstruction much simpler, causing the model to over-emphasize this objective at the expense of transcription. Insertion of an FiLM layer before the decoder degrades performance slightly. Due to the intrinsic similarities between transcription and reconstruction, this suggests that it is not helpful to augment latent vectors with detailed conditioning information in order to perform the two tasks simultaneously. Finally, training with additional audio-only data does not appear to benefit transcription under the current framework, potentially due to $L_{rc}$ being higher than $L_{tr}$, \textit{i.e.} emphasizing reconstruction over transcription.

\section{Discussion}
\label{sec:discussion}
It cannot be claimed that Timbre-Trap is superior to existing AMT frameworks, however based on empirical observations we expect that performance scales with the amount of training data and higher model complexity. We believe the framework could be further optimized with the tuning of various hyperparameters such as the pitch salience threshold or the scaling of each loss term. Our results demonstrate that the proposed framework is viable for the unification of music transcription and audio reconstruction. Our model encodes information necessary to perform both tasks simultaneously and learns how to filter out timbre in order to produce pitch salience estimates based on a simple conditioning mechanism. The degree to which this is possible validates our assumption that music transcription can be formulated as the selection of relevant frequency components during reconstruction, and therefore that transcription performance can benefit from reconstruction under this framework. With Timbre-Trap, a deeper investigation into to possibility of using partial self-supervision to learn AMT from large unlabeled audio corpuses is possible. It can also be adapted to other AMT problem domains, including note-level or instrument-aware transcription.

\begin{figure}[t]
\centering
\includegraphics[width=0.68\columnwidth]{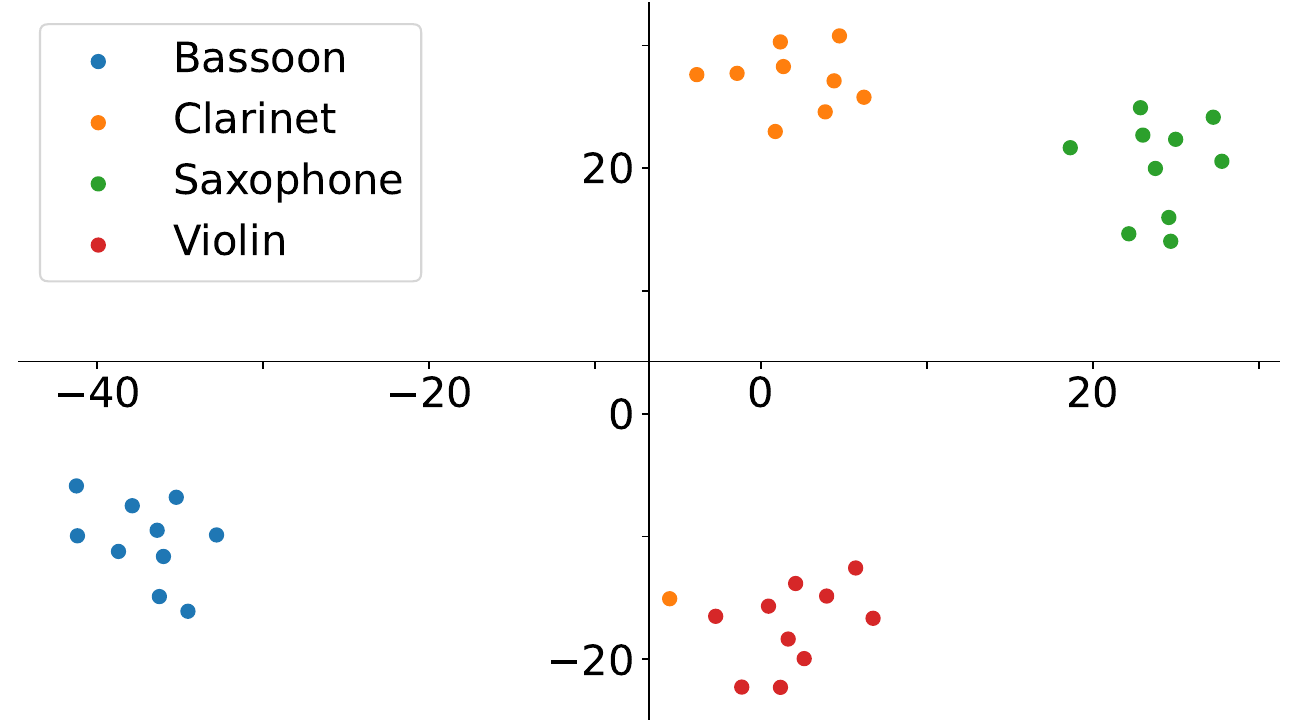}
\caption{t-SNE \cite{van2008visualizing} visualization of the average latent across each monophonic stem within Bach10, colored by associated instrument.}
\label{fig:latents}
\end{figure}

Although the use of complex CQT coefficients helps to establish a clear connection between audio and pitch salience, which is also useful in formulating $L_{cn}$, they do not appear to be essential in order to extract the benefits of reconstruction. However, complex CQT coefficients allow for the synthesis of model output, which has a few interesting implications. The Timbre-Trap formulation of transcription can be viewed as pitch-preserving style-transfer from timbre-rich audio to timbre-less audio. This means our model is capable of separating timbre information from pitch information, but it is clear from the t-distributed Stochastic Neighbor Embedding (t-SNE) \cite{van2008visualizing} latent space visualization in Figure \ref{fig:latents} that the two properties are not disentangled. Under the current framework, disentanglement is not explicitly incentivized, and therefore separation falls on the decoder. However, our framework could be adapted to carry out disentanglement by replacing or augmenting the switch mechanism with a timbre code. Score-preserving polyphonic timbre style-transfer would then consist of swapping timbre latents, and timbre-preserving polyphonic music score editing would consist of replacing pitch latents.

\section{Conclusion}
We present Timbre-Trap, a novel instrument-agnostic music transcription framework unifying transcription and audio reconstruction, where a single autoencoder is capable of performing either task via a simple switch mechanism. The framework requires little annotated data, yet can perform transcription at a level comparable to state-of-the-art methods. Moreover, invertible spectral features allow for the synthesis of coefficients corresponding to estimated pitch salience. Experiments and ablations were conducted to measure the effect of various design choices, and we suggested directions for future work.

\pagebreak
\onecolumn
\begin{multicols}{2}
\bibliographystyle{IEEEbib}
\bibliography{refs}
\end{multicols}

\end{document}